\documentclass{PoS}
\usepackage{float}
\usepackage{subfigure}

\title{WRF (Weather Research and Forecasting) model and radiative methods for cloud top height retrieval along the EUSO-SPB1 trajectory}

\ShortTitle{CTH retrieval methods for the EUSO-SPB1 experiment}

\author{S. Monte, \speaker{C. Vigorito}, M. Bertaina, S. Ferrarese, K. Shinozaki \\ 
        Universit\'a degli studi di Torino and Istituto Nazionale di Fisica Nucleare, Turin, Italy\\
        E-mail: \email{silvia.monte@edu.unito.it}}

\author{S.Briz\\
              Universidad Carlos III de Madrid, Madrid, Spain}
\author{{for the JEM-EUSO collaboration}\footnote{for collaboration list see PoS(ICRC2019)1177}}

\abstract{The Extreme Universe Space Observatory-Super Pressure Balloon (EUSO-SPB1) is a pathfinder of the JEM-EUSO program, which aims to observe Ultra High Energy Cosmic Rays (UHECRs) from near-space. It was launched from Wanaka (New Zealand) on April 25, 2017 UTC and was terminated after twelve days of flight in the South Pacific Ocean. A good knowledge of the atmospheric conditions and cloud properties, such as the Cloud Top Height (CTH), is fundamental to correctly reconstruct the energy and geometry of air showers produced by cosmic rays passing through the atmosphere. One of the methods used to retrieve the CTH is based on Numerical Weather Prediction models. In this work, we consider in particular the Weather Research and Forecasting (WRF) model. A first model test is made on the WRF parametrizations for elementary processes, applying a top-bottom directed algorithm based on two quantities: cloud fraction and optical depth. The validated procedure is then applied to the SPB1 trajectory, retrieving the CTH every ten minutes for the days of the flight. A comparison is made with the analyzed data taken from MODerate resolution Imaging Spectroradiometer (MODIS) satellite images, once per day, to understand the reliability of the method. Another way to retrieve the CTH is the so-called radiative method, that allows to calculate the Cloud Top Temperature (CTT). A vertical temperature profile is needed to transform the CTT into CTH. When radiosoundings are not available, WRF can provide vertical temperature profiles. The conversion from the CTT to the CTH is then made.}

\FullConference{36th International Cosmic Ray Conference -ICRC2019-\\
		July 24th - August 1st, 2019\\
		Madison, WI, U.S.A.}

\begin{document}

\section{Introduction}

The Extreme Universe Space Observatory-Super Pressure Balloon (EUSO-SPB1), part of the JEM-EUSO (Joint Experiment Missions) program \cite{Bertaina19}, is the first instrument aimed at detecting UHECRs Extensive Air Showers by looking down from sub-orbital space. The flight started in Wanaka, New Zealand, on April 24, 2017. This pathfinder mission was expected to fly at a constant altitude of 33 km, but it started to lose altitude due to a leak in the balloon, beginning to behave like a zero-pressure balloon and the flight needed to be terminated early, ending on May 6, 2017 in the South Pacific Ocean about 300 km SE of Easter Island. The instrument operated successfully for the duration of the flight and transmitted about 60 GB of data for 30 hours over 11 nights \cite{Wienckeicrc17}.\\
The data obtained by the JEM-EUSO instruments and their efficiency in reconstructing showers can be affected by the operational atmospheric conditions, in particular by the presence of clouds, which are active in absorbing, emitting and reflecting radiation especially in the UV range for Cherenkov and fluorescence light emitted through the entire shower development.\\
The most important parameter to evaluate the effect of clouds is the Cloud Top Height (CTH) together with the optical depth.
A space-based mission, such as the original JEM-EUSO \cite{jem1}, should constantly monitor its Field of View (FoV) by means of an Atmospheric Monitoring System (AMS) \cite{jem2}. In case of JEM-EUSO \cite{jem1} it was composed by a LIDAR and an IR camera, that, assuming a standard atmosphere (lapse rate of 0.6 K/100 m), measures the IR radiation emitted by clouds and surface allowing to retrieve the CTH. 
Different approaches are available in order to estimate the CTH. The work reported here aims at estimating the goodness of these methods, comparing the obtained results with satellite images, assumed as truth.\par
A first analysis is based on the use of Numerical Weather Prediction (NWP) models, in particular the Weather Research and Forecasting (WRF) model. After a preliminary study to examine the model sensitivity to different parameters, the cloud distribution is simulated along the entire trajectory of EUSO-SPB1 and compared, once a day, with the CTH retrieved by MODIS satellite data.\\
These simulations are later used to calculate a percentage of cloud coverage subdivided into three levels, high, medium and low clouds, and then compared to MODIS satellite data to give a general idea of the cloud distribution and to evaluate the accuracy of the algorithm.\par
A second approach is based on the Cloud Top Temperature (CTT) calculated starting from radiative equations and then converted into CTH through a vertical temperature profile provided by the WRF model. As mentioned, the IR camera assumes a standard atmosphere: NWP models can provide better information on the atmospheric conditions thereby, improving IR and LIDAR measurements. 

\section{Methods for Cloud Top Height Retrieval}
\subsection{Weather Research and Forecasting model}
\begin{figure} [H]
\centering
\includegraphics[scale=0.65, keepaspectratio]{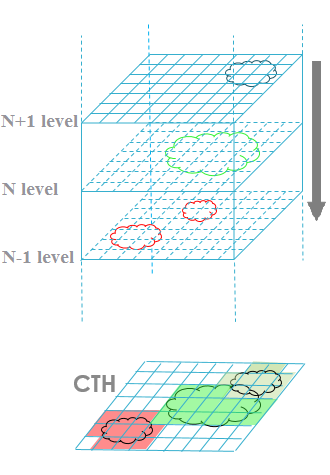}
\caption{Scheme of the algorithm for the CTH determination.}
\label{fig:algorithm}
\end{figure}

WRF is a mesoscale, fully compressible, non-hydrostatic meteorological model developed by the National Center for Atmospheric Research Mesoscale and Microscale Meteorology Division. The model uses vertical coordinates that follow the orography and it can reach up to a 1 km horizontal spatial resolution. The model uses different parametrization schemes for those processes that happen sub-grid. Version 3.9 has been used for this work. WRF provides atmospheric quantities for each pixel over all vertical levels on the model grid. Cloud fraction and optical depth, in particular, are used to evaluate the CTH, once a threshold is fixed for which the pixel is identified as a cloud. Concerning the cloud fraction \cite{Cloudfr}, the value is 0.2 while for optical depth \cite{Opticald} the value is 0.002 . Applying a top-bottom directed algorithm for each pixel, a CTH map can be reconstructed, considering only the highest level in which a "cloud" pixel is identified. A scheme of the method applied is reproduced in Figure \ref{fig:algorithm}. \\

A first sensitivity test has been made by varying the number of vertical levels, the radiation and the Planetary Boundary Layer (PBL) schemes in order to investigate the response of the model \cite{Anzalone2015}. 
The Cloud Top Pressure (CTP) retrieved is then compared to the one observed by satellite and the difference, pixel per pixel, of these two CTPs is an indication of the goodness of the model through the use of some statistical quantities, such as the median, expected to be as close as possible to zero, and the InterQuartile Range (IQR), a measure of the dispersion of the data. The best values identify the best parametrizations choice, which is now chosen to simulate the area along the EUSO-SPB1 trajectory.\par
\begin{figure}[H]
\centering
     \includegraphics[width=0.6\textwidth]{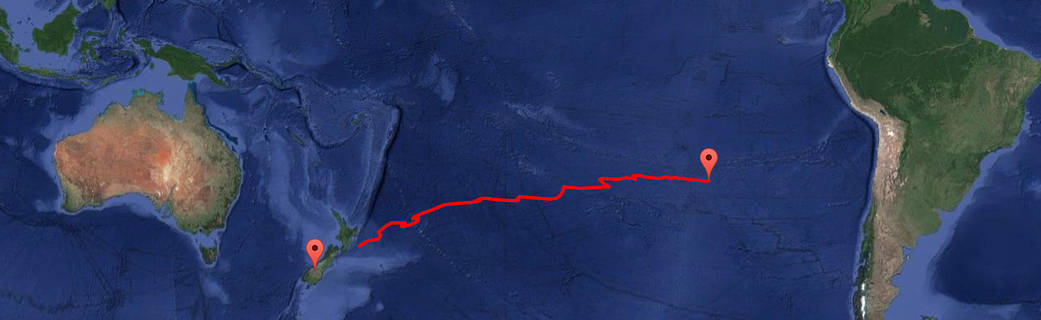} \hfill
     \includegraphics[scale=0.7, keepaspectratio]{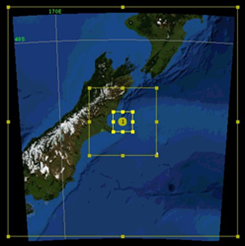}
\caption{Left: EUSO-SPB1 flight trajectory. Right: example of the horizontal nesting for April 25, 2017. }
\label{fig:domainNZ}
\end{figure}

Three nested domains (9 km, 3 km and 1 km spatial resolutions) are chosen for each day, centered on the position of the SPB1 at 12:00 UTC (left panel in Figure \ref{fig:domainNZ}) and the CTH is retrieved with WRF every ten minutes. An example of the domain selected  is reported in right panel in Figure \ref{fig:domainNZ}.\\

The obtained values are then divided into three layers: Low (LCC), Medium (MCC) and High (HCC) clouds, respectively when CTH value is between 300 m and 2 km, between 2 and 7 km and between 7 and 12 km. Two different approaches are used: first, the cloud coverage is obtained from a post-processing of the WRF model, ARW-Post \cite{Shamarock}, then the same quantity is calculated starting from the cloud fraction, as an output of the model, over all the vertical levels. 
Once per day, a satellite image along the EUSO-SPB1 trajectory is found and a comparison with WRF simulations is possible through the percentage of the cloud coverage subdivided into the three layers, selecting the same geographical area. Results will be shown in Section 3.

\subsection{Radiative Method}
\begin{figure}[H]
\centering
\includegraphics[scale=0.7,keepaspectratio]{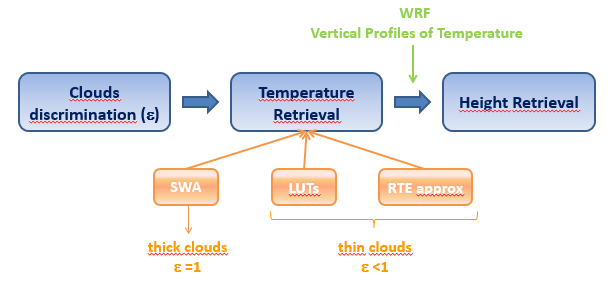}
\caption{Scheme of the radiative method: the measured emissivity allows to calculate a CTT applying different algorithms (Split-Window Algorithm -SWA-, Look Up Tables -LUTs-, Radiative Transfer Equation approximation -RTE approx-) depending on the optical depth and, using the WRF vertical profile, a CTH can be calculated.}
\label{fig:radmethod}
\end{figure}
Based on Planck's law, a correlation exists among the radiation from a cloud, its emissivity and temperature, so from measuring with an IR camera the radiance emitted by the upper cloud surface it  is possible to retrieve the CTT. Two different corrective algorithms need to be applied in the case of thin or thick clouds to take into account the atmospheric effects retrieving the real CTT and to correctly calculate the CTH, using the relationship between temperature and height provided by vertical thermal profiles \cite{Anzalone2015} (see Figure \ref{fig:radmethod}). Usually standard atmospheric profiles or observed radiosoundings are applied, but the former provides an oversimplified vertical profile while the latter are available only few times per day. \\
The WRF model can provide a vertical temperature profile for each grid point at any fixed time, allowing a more realistic CTH retrieval \cite{Tabone2015}. \\ 
In this work, the profile is obtained for every day of the flight of the EUSO-SPB1 and applied to the CTT data calculated from the cloud radiative informations. A value of CTH is found for the area seen by the IR camera.

\section{Results}
\subsection{Sensitivity test}

\begin{figure}[H]
\centering
\includegraphics[scale=0.85, keepaspectratio]{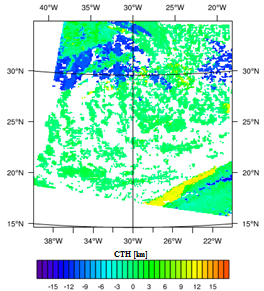}\hfill
\includegraphics[scale=0.65, keepaspectratio]{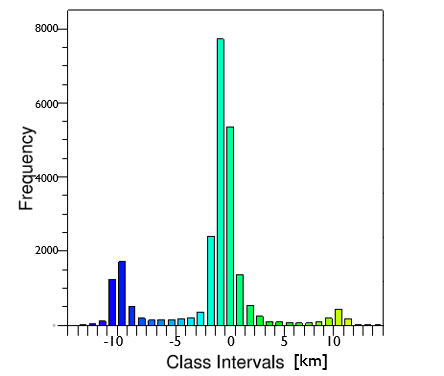}
\caption{CTH difference between MODIS and WRF and related histogram for the case reported in \cite{Anzalone2015}.}
\label{fig:difference}
\end{figure}

Prior to examining the case of  EUSO-SPB1, a fine tuning of the model is carried out to increase the performance of evaluating the cloud distribution. Different options are available for the microphysics, the radiation, the PBL and the number of vertical levels. Referring to \cite{Anzalone2015} for all the details concerning the case-study and using the cloud fraction as index, better results are obtained with 50 vertical levels, the Morrison double-moment scheme for the microphysics, the New Goddard radiation scheme and the  Milbrandt-Yamada-Nakanishi and Niino level 2.5 (MYNN 2.5) for the PBL scheme \cite{Shamarock}. The performance of the WRF model are shown in Figure \ref{fig:difference} and the Median and IQR are reported in Table \ref{table:stat17}, highlighting the improvement with respect to the previous work \cite{Anzalone2015}.

\begin{table}[H]
\centering
\begin{tabular}{|c|c|c|}
  & \textbf{Median [km]} & \textbf{IQR [km]} \\ \hline
{This work} &  -0.43 & 1.72  \\ \hline
{Previous work} & -0.85  &  10.15   \\ 
\end{tabular}
\caption{Quality of WRF model through median and IQR values for the case reported in \cite{Anzalone2015}}
\label{table:stat17}
\end{table}

The increase in quality in predicting the cloud distribution with the WRF model allows to apply these results to the case of the EUSO-SPB1, providing reliable atmospheric conditions.
 
\subsection{EUSO-SPB1 application}
The second part of the work is centered on the CTH retrieval along the EUSO-SPB1 trajectory, applying the schemes selected for the case reported in Section 3.1 . Table \ref{table:simNZ1} summarizes the central position of the WRF calculation on the left and the available MODIS images that be can compared. 

\begin{table}[H]
\begin{tabular}{ll}
\centering
\begin{tabular}{| c | c | c | c | c |}
  & \textbf{Date} & \textbf{Center point} & \textbf{Start time } & \textbf{End time} \\ 
  &        & \textbf{(Lat., Long.)}  & \textbf{(UTC)} & \textbf{(UTC)} \\ \hline
Nz1 & 24/04 & (-44.35 $^{\circ}$, +170.41 $^{\circ}$)  & 21:00  & 03:00 \\ \hline
Nz2 & 25/04 & (-43.41 $^{\circ}$, +173.56 $^{\circ}$)  &  6:00   & 18:00\\ \hline
Nz3 & 26/04 & (-40.52 $^{\circ}$, +179.20 $^{\circ}$)  & 10:00  & 17:00\\ \hline
Nz4 & 27/04 & (-38.47 $^{\circ}$, -176.95 $^{\circ}$)  & 7:00   & 17:00\\ \hline
Nz5 & 28/04 & (-37.12 $^{\circ}$, -172.41 $^{\circ}$)  & 1:00   & 18:00 \\ \hline
Nz6 & 29/04 & (-36.28 $^{\circ}$, -168.96 $^{\circ}$)  & 5:00   & 17:00\\ \hline
Nz7 & 30/04 & (-34.32 $^{\circ}$, -164.47 $^{\circ}$)  & 6:00   & 17:00 \\ \hline
Nz8 & 01/05 & (-33.03 $^{\circ}$, -156.17 $^{\circ}$)  &  3:00  & 17:00 \\ \hline
Nz9 & 02/05 & (-32.85 $^{\circ}$, -147.64 $^{\circ}$)  &  3:00  & 16:00 \\ 
\end{tabular}
\label{table:simNZ}
&
\centering
\begin{tabular}{| c | c | c |}
  & \textbf{Date} & \textbf{Time}  \\
 &   & \textbf{(UTC)} \\ \hline
Nz3 & 26/04 & 13:20\\ \hline
Nz4 & 27/04 & 12:25\\ \hline
Nz5 & 28/04 & 02:05 \\ \hline
  &       & 13:10  \\ \hline
Nz6 & 29/04 & 12:10\\ \hline
Nz7 & 30/04 & 12:50 \\ \hline
  &       & 12:55 \\ \hline
Nz8 & 01/05 & 12:00 \\ \hline
Nz9 & 02/05 & 11:05 \\ 
\end{tabular}
\end{tabular}
\caption{Left: balloon coordinates and hours for the retrieved CTH. Right: MODIS images along the EUSO-SPB1 trajectory.}
\label{table:simNZ1}
\end{table}

\begin{figure}[H]
\centering
\includegraphics[scale=0.9, keepaspectratio]{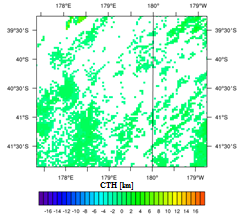}\hfill
\includegraphics[scale=0.6, keepaspectratio]{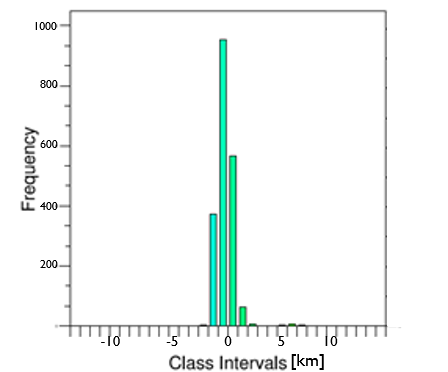}
\caption{CTH difference between MODIS and WRF and related histogram for the case of April 26, 2017 at 13:20 UTC by this work}
\label{fig:diffNZ}
\end{figure}

The data of April 26, 2017 13:20 UTC are reported here as an example (Figure \ref{fig:diffNZ}) and the Median results -0.38 km while the IQR is 1.07 km .

As an additional and more generic evaluation, MODIS- and WRF-derived CTHs are subdivided into the three defined cloud layers (Section 2.1) and a percentage of cloud coverage is calculated. The comparison is made between results obtained from ARW-Post, from the cloud fraction and from the MODIS observed CTH over the same area (Table \ref{table:percent}).

\begin{table} [H]
\centering
\begin{tabular}{| c | c | c | c | c |}
  & \textbf{HCC} & \textbf{MCC} & \textbf{LCC} & \textbf{Clear}  \\ \hline
ARWPost & 0\% & 3\% & 96\% & 0\% \\ \hline
WRF & 0\% & 2\% & 88\% & 9\% \\ \hline
MODIS & 0\% & 1\% & 59\% & 39\% \\
\end{tabular}
\caption{Percentage of cloud coverage over the three defined cloud layers for April 26, 2017.}
\label{table:percent}
\end{table}

High and Medium clouds have consistent percentages while Low clouds and Clear atmosphere differ significantly. A possible explanation is linked to the different way in which the model identifies the Low clouds. A further study is necessary.\par

\subsection{Radiative method}
\begin{figure}[H]
\centering
\includegraphics[scale=0.52,keepaspectratio]{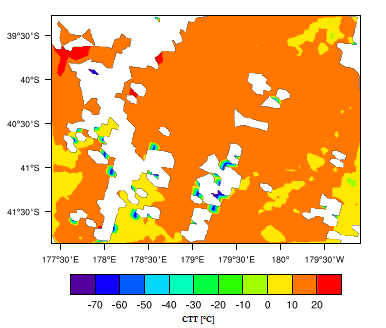} \hfill
\includegraphics[scale=0.5,keepaspectratio]{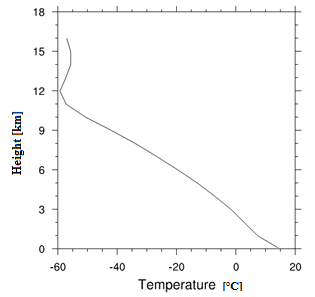} \hfill
\includegraphics[scale=0.47,keepaspectratio]{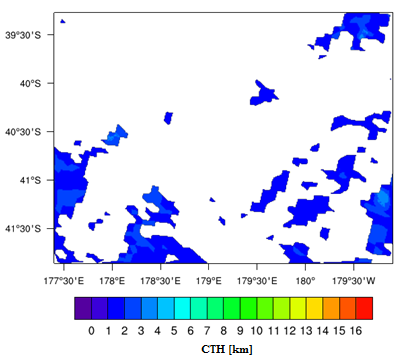}
\caption{Left: CTT calculated from the measured emissivity. Center: vertical temperature profile. Right: CTH retrieved for April 26, 2017.}
\label{fig:rad26}
\end{figure}

As explained in Section 2.2, the CTT is retrieved for April 26, 2017 in the area seen by the IR camera on board EUSO-SPB1 with a 5 km spatial resolution. A vertical temperature profile is obtained for the area with the WRF model, using the settings already described (Section 3.1). Applying this profile to the CTT is possible to obtain the CTH and the results are reported in Figure \ref{fig:rad26}.

A field of Low clouds, with values less or equal to 4 km, is obtained as it was expected from MODIS. 
 A first qualitative comparison is only possible, because a quantitative one requires a rescale of the image to a greater spatial resolution, having  the WRF simulations a 3 km resolution and the MODIS images a 1 km resolution.

\section{Conclusions}

Two methods are here investigated to retrieve the CTH along the EUSO-SPB1 trajectory in order to allow a better reconstruction of the UHECRs Extensive Air Showers and a better estimation of the exposure, which is highly affected by atmospheric conditions.\\
After a sensitivity test to identify the best parametrisations to set the WRF model, the entire trajectory is simulated every ten minutes and the CTH is obtained. The goodness of the results is evaluated for one of the days of flight pixel per pixel for each vertical level, through the median and the IQR of the difference with MODIS, and as a percentage of cloud coverage in three layers. In the former case results are good, furnishing a median value near to zero and an IQR very low. In the latter, medium and high clouds seems to be well predicted while low clouds and clear sky create some discrepancies.\par
The second method furnishes only a qualitative comparison, that seems to be reliable to the satellite data. A quantitative study needs to be made to adequately estimate the goodness of the radiative method and the use of WRF vertical temperature profiles.\par
\section{Acknowledgments}
This work was partially supported by Basic Science Interdisciplinary Research Projects of RIKEN and JSPS KAKENHI Grant (22340063,23340081, and 24244042), by the Italian Ministry of Foreign Affairs and International Cooperation, by the Italian Space Agency through the ASI INFN agreement n. 2017-8-H.0, by contract 2016-1-U.0, by NASA award 11-APRA-0058 in the USA, by the Deutsches Zentrum f\"{u}r Luft- und Raumfahrt, by the France space agency CNES, the Helmholtz Alliance for Astroparticle Physics funded by the Initiative and Networking Fund of the Helmholtz Association (Germany), and by Slovak Academy of Sciences MVTS JEMEUSO as well as VEGA grant agency project 2/0132/17. Russia is supported by ROSCOSMOS and the Russian Foundation for Basic Research Grant No 16-29-13065. Sweden is funded by the Olle Engkvist Byggm\"{a}stare Foundation.

\end{document}